\documentclass[pra,aps,twocolumn,superscriptaddress,nofootinbib]{revtex4}

\usepackage{latexsym}
\usepackage{hyperref}
\usepackage{epsfig}
\usepackage{psfrag}
\usepackage{graphics}
\usepackage{amsmath}
\usepackage{xspace}
\usepackage{amssymb}
\usepackage{amsthm}

\begin{document}

\title{Optical quantum computation using cluster states}

\author{Michael A. Nielsen} 
\email[]{nielsen@physics.uq.edu.au and www.qinfo.org/people/nielsen} 
\affiliation{School of Physical Sciences and School of Information
Technology and Electrical Engineering, \\ 
The University of Queensland, Brisbane, Queensland 4072, Australia}

\date{\today}

\begin{abstract}
  We propose an approach to optical quantum computation in which a
  deterministic entangling quantum gate may be performed using, on
  average, a few hundred coherently interacting optical elements
  (beamsplitters, phase shifters, single photon sources, and
  photodetectors with feedforward).  This scheme combines ideas from
  the optical quantum computing proposal of Knill, Laflamme and
  Milburn [Nature \textbf{409} (6816), 46 (2001)], and the abstract
  cluster-state model of quantum computation proposed by Raussendorf
  and Briegel [Phys.~Rev.~Lett. \textbf{86}, 5188 (2001)].
\end{abstract}

\maketitle

%
% section: introduction
%
Optical approaches to quantum computation are attractive due to the
long decoherence times of photons, and the relative ease with which
photons may be manipulated.  A ground breaking proposal of Knill,
Laflamme and Milburn~\cite{Knill01a} (KLM) demonstrated that
all-optical quantum computation is, in principle, possible using just
beamsplitters, phase shifters, single photon sources, and
photodetectors with feedforward.  Experimental demonstrations of
several of the basic elements of KLM have since been
performed~\cite{Pittman03a,OBrien03a,Sanaka03a}.

%
% obstacles
%
Despite these impressive successes, the obstacles to fully scalable
quantum computation with KLM remain formidable.  The biggest
challenge is to perform a two-qubit entangling gate in the
near-deterministic fashion required for scalable quantum computation.
KLM propose doing this using a combination of three ideas.  (1) Using
linear optics, single-photon sources and photodetectors,
\emph{non-deterministically} perform an entangling gate.  This gate
fails most of the time, destroying the state of the computer, and so
is not immediately suitable for quantum computation.  (2) By
combining the basic non-deterministic gate with quantum teleportation,
a class of non-deterministic gates which are not so destructive is
found.  We denote these gates $CZ_{n^2/(n+1)^2}$, where $n$ is a
positive integer.  $CZ_{n^2/(n+1)^2}$ has probability of success
$n^2/(n+1)^2$; the larger $n$ is, the greater the chance of success,
but the more complex the corresponding optical circuit.  (3) By using
quantum error-correction, the probability of the gate succeeding can
be improved until the gate is near-deterministic, allowing scalable
quantum computation.

%
% drawbacks
%
The combination of these three ideas allows quantum computation, in
principle.  Existing experimental implementations have demonstrated
(1), and promise to do (2) (for small values of $n$) in the near
future.  However, to perform $CZ_{n^2/(n+1)^2}$ for large values of
$n$, or to do step (3), is far more difficult. KLM analyse a scheme in
which the $CZ_{9/16}$ gate is combined with error-correction. To do a
single entangling gate with probability of success $95 \%$ requires
about $300$ successful $CZ_{9/16}$ gate operations, i.e., tens of
thousands of optical elements.  Higher probabilities of success
require more optical elements.

%
% what the present paper does
%
The present paper describes an approach to optical quantum computation
that makes use of ideas (1) and (2) (for $n =1$ and $n=2$), but avoids
step (3).  The scheme combines KLM's non-deterministic gates with the
\emph{cluster-state} model of quantum computation proposed by
Raussendorf and Briegel~\cite{Raussendorf01a}. Using a $CZ_{4/9}$ gate
(which uses roughly $2-3$ times fewer optical elements than the
$CZ_{9/16}$ gate ) a single logical quantum gate in this proposal
requires, on average, fewer than $8$ successful $CZ_{4/9}$ gates.  In
this scheme there is an additional overhead due to the single-qubit
gates; even when that is taken into account, fewer than $24$
$CZ_{4/9}$ gates are required to simulate an entangling gate.  This is
not only substantially simpler than KLM, but the resulting logical
gates work \emph{deterministically} (assuming ideal optical elements),
as opposed to the $5 \%$ error experienced by KLM's entangling gates.

%
% Yoran and Reznik
%
Yoran and Reznik~\cite{Yoran03a} have proposed a scheme for optical
quantum computation based on KLM, but using substantially simpler
resources.  This scheme has several elements in common with the
current proposal, including offline preparation of a quantum state,
which is used to do computation deterministically.  (These
similarities bear further investigation; although~\cite{Yoran03a} does
not use the cluster-state model of computation, their method has many
similarities.)  For comparison,~\cite{Yoran03a} estimate $20-30$
$CZ_{9/16}$ gates per logical gate, or perhaps $2-3$ times as many
optical elements as the cluster-state proposal, due to the greater
complexity of the $CZ_{9/16}$ gate.

%
% section: the cluster-state model
%
\textbf{Cluster-state quantum computation:} The cluster-state model of
quantum computation~\cite{Raussendorf01a} is a beautiful alternate
model of quantum computation, mathematically equivalent to the
standard quantum circuit model, but quite different in physical
aspect. We describe briefly the procedure used to simulate a quantum
circuit in the cluster-state model; proofs may be found
in~\cite{Raussendorf01a}.  Note that this is an abstract model for
quantum computation, not a proposal for physical implementation, so we
describe it without reference to the details of a specific physical
system.

%
% example
%
To simulate a quantum circuit like that in
Fig.~\ref{fig:basic-circuit} we first prepare the \emph{cluster
  state}, an entangled network of qubits defined as in
Fig.~\ref{fig:simulating-cluster-state}.  Each qubit in the quantum
circuit is replaced by a horizontal line of qubits in the cluster
state.  Different horizontal qubits represent the original qubit at
different times, with the progress of time being from left to right.
Each single-qubit gate in the quantum circuit is replaced by two
horizontally adjacent qubits in the cluster state.  (Alternately, one
or three horizontally adjacent qubits could be used; that would
correspond to slightly different classes of single-qubit unitaries
being simulated.)  {\sc cphase} gates in the original circuit are
simulated using a vertical ``bridge'' connecting the appropriate
qubits.

\newsavebox{\cphase}
\savebox{\cphase}(0,0){ \setlength{\unitlength}{1cm}
 \put(0,1){\line(0,-1){1.0}}              % line in c-phase
 \put(0,0){\circle*{0.2}}                 % top end of c-phase
 \put(0,1){\circle*{0.2}}                 % bottom end of c-phase
}

\begin{figure}[ht]
\begin{center}
\setlength{\unitlength}{1cm}
\begin{picture}(5.2,1.4)
\put(0,1.05){$|+\rangle$}                  % input to line 1
\put(0,0.05){$|+\rangle$}                  % input to line 2
\put(0.5,1.1){\line(1,0){0.5}}               % first part of line 1
\put(0.5,0.1){\line(1,0){0.5}}               % first part of line 2
\put(1.0,0.7){\framebox(1.0,0.8){$U_{\alpha_1,\alpha_2}$}}              
           % first unitary
\put(1.0,-0.3){\framebox(1.0,0.8){$U_{\beta_1,\beta_2}$}}
           % second unitary
\put(2.0,1.1){\line(1,0){1.6}}               % line 1
\put(2.0,0.1){\line(1,0){1.6}}               % line 2
\put(2.8,0.6){\usebox{\cphase}}            % c-phae gate
\put(3.6,0.7){\framebox(1.0,0.8){$U_{\alpha_3,\alpha_4}$}}              
           % third unitary
\put(3.6,-0.3){\framebox(1.0,0.8){$U_{\beta_3,\beta_4}$}}
           % fourth unitary
\put(4.6,1.1){\line(1,0){0.5}}               % line 1
\put(4.6,0.1){\line(1,0){0.5}}               % line 2
\end{picture}
\end{center}
\caption{A two-qubit quantum circuit. Without loss of generality we
  assume the computation starts in the $|+\rangle \equiv
  (|0\rangle+|1\rangle)/\sqrt{2}$ state, since single-qubit gates can
  be prepended to the circuit if we wish to start in some other state.
  The boxes are single-qubit gates, $U_{\alpha,\alpha'} \equiv
  X_{\alpha'} Z_{\alpha}$, denoting a rotation by $\alpha$ about the
  $\hat z$ axis of the Bloch sphere, followed by a rotation by
  $\alpha'$ about the $\hat x$ axis.  The two-qubit gate is a
  controlled-phase ({\sc cphase}) gate, whose action in the
  computational basis is $|ab\rangle \rightarrow (-1)^{ab}
  |ab\rangle$. {\sc cphase} and the single-qubit operations
  $U_{\alpha, \alpha'}$ are together universal for quantum
  computation.\label{fig:basic-circuit}}
\end{figure}
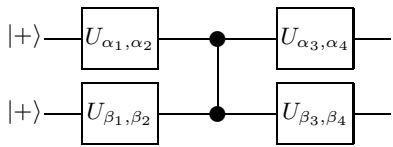

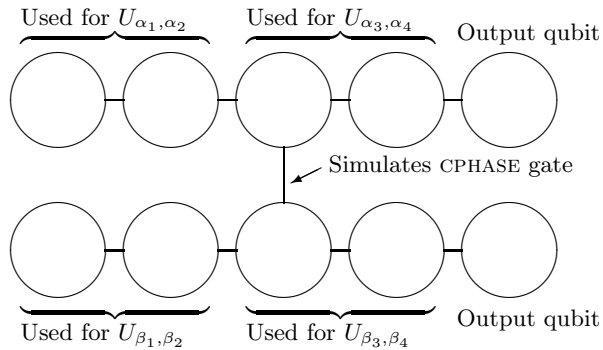
\begin{figure}[h]
\begin{center}
\setlength{\unitlength}{1cm}
\begin{picture}(6.1,3)
\put(-0.5,2.9){$\overbrace{\hspace{2.5cm}}$}    % qubits 1 and 2, top line
\put(-0.5,3.2){Used for $U_{\alpha_1,\alpha_2}$}
\put(2.5,2.9){$\overbrace{\hspace{2.5cm}}$}     % qubit 3 and 4, top line
\put(2.5,3.2){Used for $U_{\alpha_3,\alpha_4}$}
\put(-0.5,-0.5){$\underbrace{\hspace{2.5cm}}$}  % qubits 1 and 2, bot line 
\put(-0.5,-1.0){Used for $U_{\beta_1,\beta_2}$}
\put(2.5,-0.5){$\underbrace{\hspace{2.5cm}}$}   % qubits 3 and 4, bot line 
\put(2.5,-1.0){Used for $U_{\beta_3,\beta_4}$}
\put(5.3,3.0){Output qubit}   % top output
\put(5.3,-0.8){Output qubit}   % bottom output

\put(3.5,1.3){\vector(-2,-1){0.4}}  % label on verticle bridge
\put(3.6,1.25){Simulates {\sc cphase} gate}  

\put(0.0,2.2){\circle{1.2}}       % qubit 1, top line
\put(0.62,2.2){\line(1,0){0.26}}
\put(0.0,0.2){\circle{1.2}}       % qubit 1, bottom line
\put(0.62,0.2){\line(1,0){0.26}}
\put(1.5,2.2){\circle{1.2}}       % qubit 2, top line
\put(2.12,2.2){\line(1,0){0.26}}  
\put(1.5,0.2){\circle{1.2}}       % qubit 2, bottom line             
\put(2.12,0.2){\line(1,0){0.26}}
\put(3.0,2.2){\circle{1.2}}       % qubit 3, top line
\put(3.62,2.2){\line(1,0){0.26}}  
\put(3.0,0.2){\circle{1.2}}       % qubit 3, bottom line             
\put(3.62,0.2){\line(1,0){0.26}}
\put(3.0,0.82){\line(0,1){0.76}}  % vertical bridge
\put(4.5,2.2){\circle{1.2}}       % qubit 4, top line
\put(5.12,2.2){\line(1,0){0.26}}  
\put(4.5,0.2){\circle{1.2}}       % qubit 4, bottom line             
\put(5.12,0.2){\line(1,0){0.26}}
\put(6.0,2.2){\circle{1.2}}       % qubit 5, top line
\put(6.0,0.2){\circle{1.2}}       % qubit 5, bottom line             
\end{picture}
\end{center}
\vspace{0.5cm}
\caption{The cluster state used to simulate the circuit in 
  Fig.~\ref{fig:basic-circuit}. Each circle represents a single qubit.
  The cluster state is constructed by preparing each qubit in the
  state $|+\rangle \equiv (|0\rangle+|1\rangle)/\sqrt 2$, and then
  applying {\sc cphase} between any two qubits joined by a
  line.  Since the {\sc cphase} gates commute with one another, it
  does not matter in what order they are applied.
\label{fig:simulating-cluster-state}}
\end{figure}

\begin{figure}[ht]
\begin{center}
\setlength{\unitlength}{1cm}
\begin{picture}(6.1,2.6)
\put(-0.5,2.1){$1. H Z_{\alpha_1}$}   % qubit 1, top line
\put(-0.5,0.1){$1. H Z_{\beta_1}$}   % qubit 1, bot line
\put(1.0,2.1){$2. H Z_{\alpha_2'}$}   % qubit 2, top line
\put(1.0,0.1){$2. H Z_{\beta_2'}$}   % qubit 2, bot line
\put(2.5,2.1){$3. H Z_{\alpha_3'}$}   % qubit 3, top line
\put(2.5,0.1){$3. H Z_{\beta_3'}$}   % qubit 3, bot line
\put(4.0,2.1){$4. H Z_{\alpha_4'}$}   % qubit 4, top line
\put(4.0,0.1){$4. H Z_{\beta_4'}$}   % qubit 4, bot line

\put(0.0,2.2){\circle{1.2}}       % qubit 1, top line
\put(0.62,2.2){\line(1,0){0.26}}
\put(0.0,0.2){\circle{1.2}}       % qubit 1, bottom line
\put(0.62,0.2){\line(1,0){0.26}}
\put(1.5,2.2){\circle{1.2}}       % qubit 2, top line
\put(2.12,2.2){\line(1,0){0.26}}  
\put(1.5,0.2){\circle{1.2}}       % qubit 2, bottom line             
\put(2.12,0.2){\line(1,0){0.26}}
\put(3.0,2.2){\circle{1.2}}       % qubit 3, top line
\put(3.62,2.2){\line(1,0){0.26}}  
\put(3.0,0.2){\circle{1.2}}       % qubit 3, bottom line             
\put(3.62,0.2){\line(1,0){0.26}}
\put(3.0,0.82){\line(0,1){0.76}}  % vertical bridge
\put(4.5,2.2){\circle{1.2}}       % qubit 4, top line
\put(5.12,2.2){\line(1,0){0.26}}  
\put(4.5,0.2){\circle{1.2}}       % qubit 4, bottom line             
\put(5.12,0.2){\line(1,0){0.26}}
\put(6.0,2.2){\circle{1.2}}       % qubit 5, top line
\put(6.0,0.2){\circle{1.2}}       % qubit 5, bottom line             
\end{picture}
\end{center}
\caption{The circuit in 
  Fig.~\ref{fig:basic-circuit} is simulated by measuring the
  individual qubits of the cluster in the time order denoted by the
  labels on the qubits, $1,2,3,\ldots$.  Note that some qubits have
  the same label, and may be measured in either order, or
  simultaneously.  The measurement basis is indicated via a
  single-qubit unitary operation to be applied before measuring in the
  computational basis.  For example, the first qubit on the top line
  has $HZ_{\alpha_1}$ applied, before measuring in the computational
  basis.  The prime notation, e.g.  $\alpha_3'$, indicates that the
  value of $\alpha_3'$ is either $\pm \alpha_3$, with the sign
  determined by the outcome of previous measurements, as described
  in~\cite{Raussendorf01a}.
\label{fig:simulating-cluster-state-2}}
\end{figure}
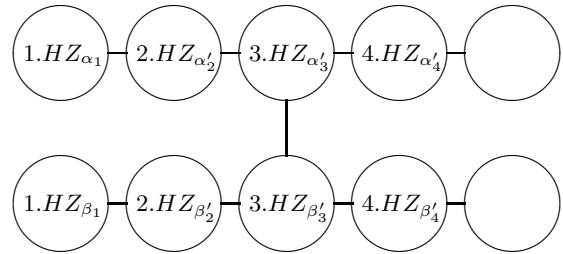

%
% how to do a computation
%
With the cluster state prepared, simulation of the circuit is
accomplished using single-qubit measurements, and feedforward of
measurement results to control the basis in which later measurements
are performed.  The sequence of measurements is illustrated in
Fig.~\ref{fig:simulating-cluster-state-2}.  The output of the circuit
in Fig.~\ref{fig:basic-circuit} is the same as the state of the qubits
at the end of the horizontal lines in
Fig.~\ref{fig:simulating-cluster-state-2}, up to a known product of
Pauli matrices, which can be compensated for by classical
postprocessing.  Extending this example along similar lines, we can
simulate an arbitrary quantum circuit using just cluster-state
preparation, single-qubit unitaries, measurements in the computational
basis, and measurement feedforward~\cite{Raussendorf01a}.  

%
% my simplifications
%
For convenience we have presented the cluster-state model in a
slightly different form than~\cite{Raussendorf01a}.
In~\cite{Raussendorf01a} the vertical bridges contain two additional
intermediate qubits in order to simulate a {\sc cphase} gate.  The
reason~\cite{Raussendorf01a} has this more complicated bridge is
because they assume that the quantum circuit one wishes to simulate is
not known until \emph{after} preparation of the cluster state.
\cite{Raussendorf01a} make this assumption in order to show that a
\emph{single} cluster state can simulate an \emph{arbitrary} quantum
computation of a given breadth and depth. In implementation one knows
the circuit beforehand (e.g. Shor's circuit for
factoring~\cite{Shor97a}), and the intermediate qubits in the bridge
can be dispensed with.  The simpler bridge, while not essential, does
simplify the optical implementation.

%
% discarding
%
To combine the cluster-state model with KLM we need one final
observation about the properties of cluster states.  Using the
definition of the cluster state and the {\sc cphase} gate we obtain
the following expression for the cluster state, up to normalization,
\begin{eqnarray} \label{eq:expression-cluster-state}
  \sum_{z_1,z_2,\ldots} (-1)^{\sum_{j,k} z_j z_k} |z_1,z_2,\ldots\rangle,
\end{eqnarray}
where the first sum is over all configurations $z_1,z_2,\ldots$ $(z_j
= 0,1)$ of the qubits making up the cluster state, and the sum in the
exponent is over all pairs $(j,k)$ of neighbouring qubits in the
cluster.  Suppose we measure one of the cluster qubits in the
computational basis, with outcome $m$.  It follows from
Eq.~(\ref{eq:expression-cluster-state}) that the posterior state is
just a cluster state with that node deleted, up to a local $Z^m$
operation applied to each qubit neighbouring the deleted qubit.  These
are known local unitaries, whose effect may be compensated in
subsequent operations, so we may effectively regard such a
computational basis measurement as simply removing the qubit from the
cluster.

%
% section
%
\textbf{KLM optical quantum computation:} KLM encodes a single qubit
in two optical modes, $A$ and $B$, with logical qubit states
$|0\rangle_L \equiv |01\rangle_{AB}$, and $|1\rangle_L \equiv
|10\rangle_{AB}$.  State preparation is done using single-photon
sources, while measurements in the computational basis may be achieved
using high-efficiency photodetectors.  Such sources and detectors make
heavy demands not entirely met by existing optical technology,
although encouraging progress on both fronts has been reported
recently.  Arbitrary single-qubit operations are achieved using phase
shifters and beamsplitters.

%
% the non-det dest gate
%
The main difficulty in KLM is achieving near-deterministic entangling
interactions between qubits.  KLM propose two basic constructions, one
building upon the other, for implementing a \emph{non-deterministic}
{\sc cphase} gate, that is, a gate which with some probability
succeeds, and with some probability fails, and whether the gate
succeeds or fails is known.  The two constructions differ in their
success probability, and in whether failure results in the destruction
of the qubits, or in some incorrect operation being applied.  We now
summarize the basic properties of the two constructions.

\emph{The destructive non-deterministic {\sc cphase} gate:} We
describe a construction of Knill~\cite{Knill02c} that slightly
simplifies the original KLM construction.  Knill's construction takes
two logical qubits as input, and with probability $2/27$ applies a
{\sc cphase} gate, or else fails, destroying the state of the qubits.
The gate uses two phase shifters, four beamsplitters, two
single-photon ancillas, and two photodetectors measuring the ancillas;
these must be capable of distinguishing $1$ photon from $0$ or $2$
photons.

%
% improved gate
%
\emph{Non-destructive non-deterministic {\sc cphase} gates:} The gate
just described can be improved by combining it with the idea of gate
teleportation~\cite{Gottesman99a,Nielsen97c,Bennett93a}.  The result
is a gate $CZ_{n^2/(n+1)^2}$ which with probability $n^2/(n+1)^2$
applies a {\sc cphase} to two input qubits, where $n$ is a positive
integer.  When the gate fails, the effect is to perform a measurement
of those qubits in the computational basis.  Increasing values of $n$
correspond to increasingly complicated teleportation circuits.  The
only two values of $n$ we shall need are $n=1$ and $n=2$, both of
which use relatively simple teleportation circuits, with just a few
optical elements --- for $n=1$, $8$ beamsplitters, $4$ photodetectors,
and $4$ single-photon preparations; for $n=2$ less than $70$
beamsplitters, $30$ photodetectors, and $12$ single-photon
preparations.

%
% what happens when we sequentially apply the gate
%
The basic $CZ_{n^2/(n+1)^2}$ gate involves two teleportation steps
performed in parallel on the two qubits, succeeding with independent
probabilities $n/(n+1)$.  It is possible to perform these
teleportations sequentially, with the result~\cite{Knill01a} that if
the first teleportation fails, we can abort the gate, without harming
the second qubit.  More generally, if we wish to perform {\sc cphase}
gates between a single qubit $S$, and several other qubits $A,
B,\ldots$, it is possible to first perform all the teleportation steps
involving just qubit $S$, and abort if any fail, preserving qubits
$A,B,\ldots$.  If they all succeed, the remaining teleportation steps
involving the other qubits are performed, each with probability of
success $n/(n+1)$.  Doing the gates in this sequential way will have
considerable advantages in the cluster-state model of quantum
computation.

%
% what else KLM
%
KLM achieves scalable quantum computation by combining quantum
error-correction and the elements we have described to develop a {\sc
  cphase} gate that succeeds with much higher probability.  This
construction is avoided in the cluster-state implementation of optical
quantum computation, and so we omit a description.

%
% section: combined scheme
%
\textbf{Optical quantum computation with cluster states:} The idea is
to build up the cluster state by non-deterministically adding extra
qubits to the cluster using $CZ_{4/9}$ or $CZ_{1/4}$ gates.  If this
can be done, the other operations in the cluster-state model can be
done following KLM's prescription.  To simplify preparation it helps
to suppose that each qubit in the cluster is involved in at most a
\emph{single} vertical bridge.  The only reason more vertical
connections might be required is if the quantum circuit being
simulated involves the same qubit in multiple parallel {\sc cphase}
gates.  We may assume this does not occur, without affecting the
ability of a cluster-state computation to efficiently simulate a
quantum circuit.

%
% how we build the cluster up
%
We will build the cluster up by interleaving two types of operation:
attempting to add a site connected to the current cluster through a
single bond, and attempting to add a site connected to the current
cluster through a double bond.  It is not difficult to see that any
cluster can be built up by alternating operations of this type.  We
analyse the two cases separately.

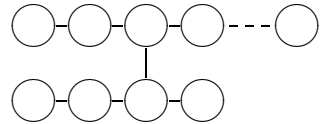
\begin{figure}[ht]
\begin{center}
\setlength{\unitlength}{0.5cm}
\begin{picture}(7.5,2.0)
\put(0.0,1.7){\circle{1.2}}       % qubit 1, top line
\put(0.62,1.7){\line(1,0){0.26}}
\put(0.0,-0.3){\circle{1.2}}       % qubit 1, bottom line
\put(0.62,-0.3){\line(1,0){0.26}}
\put(1.5,1.7){\circle{1.2}}       % qubit 2, top line
\put(2.12,1.7){\line(1,0){0.26}}  
\put(1.5,-0.3){\circle{1.2}}       % qubit 2, bottom line             
\put(2.12,-0.3){\line(1,0){0.26}}
\put(3.0,1.7){\circle{1.2}}       % qubit 3, top line
\put(3.62,1.7){\line(1,0){0.26}}  
\put(3.0,-0.3){\circle{1.2}}       % qubit 3, bottom line             
\put(3.62,-0.3){\line(1,0){0.26}}
\put(3.0,0.32){\line(0,1){0.76}}  % vertical bridge
\put(4.5,1.7){\circle{1.2}}       % qubit 4, top line
\put(4.5,-0.3){\circle{1.2}}       % qubit 4, bottom line             

\put(7.0,1.7){\circle{1.2}}       % qubit 5, top line
\multiput(5.22,1.7)(0.42,0){3}{\line(1,0){0.21}} 
\end{picture}
\end{center}
\vspace{-0.1cm}
\caption{Attempting to add a site connected by a single bond to the current
  cluster, using a $CZ_{4/9}$ gate.  By performing the gate with
  sequential teleportations we ensure that the probability of success
  is $2/3$.
\label{fig:add-single-bond}}
\end{figure}

%
% adding a single-bond site
%
The procedure to add a site connected by a single single bond to the
cluster is illustrated in Fig.~\ref{fig:add-single-bond}.  With
probability $2/3$ this succeeds, and a site is added to the cluster,
while with probability $1/3$ it fails, and a measurement in the
computational basis removes a qubit from the cluster, namely, the
qubit with which a {\sc cphase} was attempted.  Thus, the expected
number of sites added to the cluster is $2/3 \times 1 + 1/3 \times
(-1) = 1/3$.

\begin{figure}[ht]
\begin{center}
\setlength{\unitlength}{0.5cm}
\begin{picture}(7.5,2.0)
\put(0.0,1.7){\circle{1.2}}       % qubit 1, top line
\put(0.62,1.7){\line(1,0){0.26}}
\put(0.0,-0.3){\circle{1.2}}       % qubit 1, bottom line
\put(0.62,-0.3){\line(1,0){0.26}}
\put(1.5,1.7){\circle{1.2}}       % qubit 2, top line
\put(2.12,1.7){\line(1,0){0.26}}  
\put(1.5,-0.3){\circle{1.2}}       % qubit 2, bottom line             
\put(2.12,-0.3){\line(1,0){0.26}}
\put(3.0,1.7){\circle{1.2}}       % qubit 3, top line
\put(3.62,1.7){\line(1,0){0.26}}  
\put(3.0,-0.3){\circle{1.2}}       % qubit 3, bottom line             
\put(3.62,-0.3){\line(1,0){0.26}}
\put(3.0,0.32){\line(0,1){0.76}}  % vertical bridge
\put(4.5,1.7){\circle{1.2}}       % qubit 4, top line
\put(4.5,-0.3){\circle{1.2}}       % qubit 4, bottom line             
\put(4.3,-0.5){$B$}

\put(7.0,1.7){\circle{1.2}}       % qubit 5, top line
\put(6.8,1.5){$A$}
\put(5.12,1.7){\line(1,0){1.3}} 
\put(7.0,-0.3){\circle{1.2}}       % qubit 5, bot line
\put(6.8,-0.5){$S$}
\multiput(5.22,-0.3)(0.42,0){3}{\line(1,0){0.21}} 
\multiput(7.0,0.38)(0,0.26){3}{\line(0,1){0.13}} 

\end{picture}
\end{center}
\vspace{-0.1cm}
\caption{Adding a site, $S$, to the cluster by attaching it to
  two qubits $A$ and $B$ already in the cluster.
\label{fig:add-double-bond}}
\end{figure}
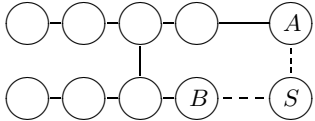

%
% adding a double-bond site
%
The procedure used to add a site connected to the current cluster by a
double bond is illustrated in Fig.~\ref{fig:add-double-bond}.  We
sequentially attempt $CZ_{4/9}$ gates between qubits $S$ and $A$, and
$S$ and $B$.  As described earlier, this can be done so that each gate
works with probability $2/3$.  If the gate between $S$ and $A$ fails,
then qubit $A$ is removed from the cluster, and we abort the
procedure.  This occurs with probability $1/3$.  If it succeeds, then
we attempt $CZ_{4/9}$ between qubit $S$ and $B$.  If this fails, then
qubit $B$ is removed from the cluster, and we abort the procedure.
This occurs with probability $2/9$.  If both gates succeed then we add
qubit $S$ to the cluster.  This occurs with probability $4/9$.  The
expected number of sites added to the cluster is thus $-1/9$.

%
% now do the random walk analysis
%
Observe that any cluster may be built up by alternating two steps: (a)
attempting to add \emph{one or more} sites that are connected to the
current cluster by a single bond, and (b) attempting to add \emph{just
  one} site that is connected by a double bond.  We conclude that for
every two attempts to add a site, the average number of sites added to
the cluster is at least $1/3 -1/9 = 2/9$.  Thus a cluster of size
$s(n)$ can be grown using roughly $9 s(n)$ attempts to add a single
site.  For maximal space efficiency, note that if the circuit being
simulated has breadth $n$ and depth $d(n)$, then it suffices to
prepare only the $O(n \log(d(n)))$ leftmost qubits in the cluster,
adding extra qubits to the cluster as earlier qubits are measured,
without danger of destroying the cluster.

%
% resource requirements
%
What resources are required to simulate a standard quantum circuit, in
this proposal?  If we assume that the circuit being simulated involves
two single-qubit unitaries $U_{\alpha,\alpha'}$ for each {\sc cphase}
gate (one on each qubit), then for every three gates we will need to
add four sites to the cluster, which means less than $24$ successful
$CZ_{4/9}$ gates.  That is, fewer than $8$ successful $CZ_{4/9}$ gates
are used per logical gate.  Note, however, that both the single-qubit
and entangling gates in the original quantum circuit require these
$CZ_{4/9}$ gates, so it is perhaps fairest to use an estimate of about
$24$ successful $CZ_{4/9}$ gates per entangling gate.  Even with this
caveat, these requirements are quite modest compared with other
proposals.

%
% alternate model
%
We have described a scheme for optically preparing cluster states and
using them for computation.  Many alternate approaches to preparation
may be conceived.  One interesting approach is illustrated in
Fig.~\ref{fig:microclusters}.  ``Microclusters'' are
non-deterministically prepared and then ``glued'' together using
$CZ_{1/4}$ gates, in order to create the cluster.  An advantage of
this approach is that the basic elements are $CZ_{1/4}$ gates, instead
of the more complex $CZ_{4/9}$ gates.  In the short term this is
likely to be simpler to implement, and to offer proof-of-principle
experimental demonstrations.  Over the long run, however, the
polynomial overhead incurred by this scheme means that the scheme
based on $CZ_{4/9}$ gates is more promising.

\newsavebox{\microcluster}
\savebox{\microcluster}(0,0){
\setlength{\unitlength}{0.5cm}
\put(0.0,1.4){\circle{0.8}}       % central node
\put(0.3,1.7){\line(3,2){0.82}}
\put(0.4,1.4){\line(1,0){0.7}}
\put(0.3,1.1){\line(3,-2){0.82}}
\put(1.5,2.4){\circle{0.8}}       % top dangling lead
\put(1.5,1.4){\circle{0.8}}       % middle dangling lead
\put(1.5,0.4){\circle{0.8}}       % bottom dangling lead
}

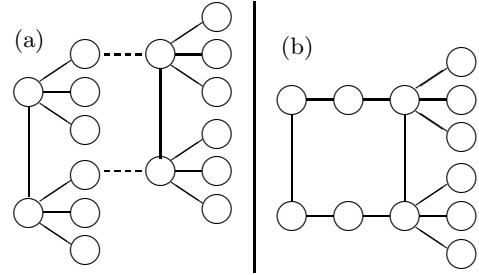
\begin{figure}[ht]
\begin{center}
\setlength{\unitlength}{0.5cm}
\begin{picture}(12,6.9)

\put(-0.4,5.5){(a)}

\put(0,4.1){\usebox{\microcluster}}
\put(0,0.9){\usebox{\microcluster}}
\put(0,1.53){\line(0,1){2.37}}

\put(3.5,5.1){\usebox{\microcluster}}
\put(3.5,2.0){\usebox{\microcluster}}
\put(3.5,2.53){\line(0,1){2.37}}

\multiput(2.0,5.3)(0.28,0){4}{\line(1,0){0.14}}
\multiput(2.0,2.2)(0.28,0){4}{\line(1,0){0.14}}

%
% second half of figure
%
\thicklines
\put(6.0,-0.5){\line(0,1){7.2}}
\thinlines

\put(6.7,5.3){(b)}
\put(7.0,1.0){\circle{0.8}}       % qubit 1, bot
\put(7.4,1.0){\line(1,0){0.7}}
\put(8.5,1.0){\circle{0.8}}       % qubit 2, bot
\put(8.9,1.0){\line(1,0){0.7}}
\put(10.0,0.8){\usebox{\microcluster}}

\put(7.0,4.1){\circle{0.8}}       % qubit 1, top
\put(7.4,4.1){\line(1,0){0.7}}
\put(8.5,4.1){\circle{0.8}}       % qubit 2, top
\put(8.9,4.1){\line(1,0){0.7}}
\put(10.0,3.9){\usebox{\microcluster}}

\put(7.0,1.4){\line(0,1){2.3}}    % first vertical bridge
\put(10.0,1.4){\line(0,1){2.3}}    % second vertical bridge

\end{picture}
\end{center}
\vspace{-0.3cm}
\caption{(a) We prepare microclusters non-deterministically, 
  as illustrated, and glue them together using parallel $CZ_{1/4}$
  gates to give a cluster like that shown in (b).  If gluing fails, we
  discard the qubits from the cluster.  The extra dangling nodes
  enable multiple attempts at adjoining a microcluster; by increasing
  the number of dangling nodes we can increase the probability of
  successful gluing.  For a cluster of size $s(n)$, using
  microclusters with $O(\log(s(n)))$ dangling nodes gives a high
  probability of successfully preparing the entire cluster.  The
  disadvantage is that preparing the microclusters
  non-deterministically incurs a polynomial overhead.
\label{fig:microclusters}}
\end{figure}

\textbf{Conclusion:} By combining the abstract cluster-state model of
computation with KLM we obtain a scheme for optical quantum
computation significantly less demanding than in existing schemes
based on single-photon preparation, linear optics, and photodetectors.
How it compares with schemes using different basic elements, like the
coherent-state scheme of Ralph \emph{et al}~\cite{Ralph03a}, depends
on future technological developments.  Work is underway to simplify
the scheme further, and to address the question of fault-tolerance.

\acknowledgments

Especial thanks to Alexei Gilchrist, whose careful reading located a
significant error in my original formulation.  Thanks also to Chris
Dawson, Jennifer Dodd, Andrew Doherty, Tim Ralph, Terry Rudolph and
Andrew White for helpful discussions.

%\bibliography{../../mybib}

\end{document}